\documentstyle[sprocl,axodraw]{article}
\def\bcn{\begin{center}}
\def\ecn{\end{center}}
\newcommand{\tenrm}{\mbox{}}
\newcommand{\bea}{\begin{eqnarray}}
\newcommand{\eea}{\end{eqnarray}}
\newcommand{\beq}{\begin{equation}}
\newcommand{\eeq}{\end{equation}}

\newcommand{\sm}{standard model}

\newcommand{\xs}{cross section}
\newcommand{\br}{branching ratio}

\newcommand{\EW}{electroweak}

\newcommand{\cm}{center of mass}

\newcommand{\ep}{\mbox{$e^+e^-$}}
\newcommand{\ee}{\mbox{$e^-e^-$}}

\def\lr3{$SU(3)_L\otimes SU(3)_R$}

\def\z0{$Z^0$}
\def\Z0{$Z^0$}

\def\ep{$e^+e^-$}

\def\cm{centre of mass}

\def\gsim{\buildrel{\lower.7ex\hbox{$>$}}\over{\lower.7ex\hbox{$\sim$}}}
\def\lsim{\buildrel{\lower.7ex\hbox{$<$}}\over{\lower.7ex\hbox{$\sim$}}}

\def\be{\begin{equation}}
\def\ee{\end{equation}}
\def\bea{\begin{eqnarray}}
\def\eea{\end{eqnarray}}

\begin{document}

\title{MEASURING TOP QUARK ELECTROWEAK DIPOLE MOMENTS}

\author{Frank Cuypers}

\address{{\tt cuypers@mppmu.mpg.de}\\
        Max-Planck-Institut f\"ur Physik,
        Werner-Heisenberg-Institut,
        F\"ohringer Ring 6,
        D--80805 M\"unchen,
        Germany}


\maketitle\abstracts{
We analyze the pair-production of top quarks
in polarized \ep\ scattering
in the presence of \EW\ dipole moments.
For this,
we consider two $CP$-odd observables
which probe both the real and the imaginary parts of these moments.
Polarizing the electron beam
turns out to be a main asset.
}

\section{Introduction}

With its extreme mass
the top quark assumes a very particular role
in the \sm\ zoology.
Because its weak decay takes place before it can hadronize~\cite{bdkkz},
the top can be studied in a much cleaner way
than any other quark.
Moreover,
since in all theories involving $CP$ violation
(including the \sm)
\EW\ dipole moments of fermions
are expected to be proportional to their mass,
the top quark is a privileged candidate for observing
such dipole moments.
A vast literature has already been devoted to this subject~\cite{pr1}.
We present here a short analysis
devoted to the measurement of
\EW\ dipole moments of the top quark
in high energy $e^+e^-$ collisions.
A presentation of similar material
has been published with S.D. Rindani~\cite{cr}.

In the following
we write all our results
in the limit where the weak mixing angle $\theta_w$
is such that
$\sin^2\theta_w=1/4$.
This is a good approximation
which greatly simplifies most analytic expressions,
because the vector coupling of the $Z^0$ to electrons then vanishes.
The unapproximated formulas are given in Ref.~\cite{cr}
and all our numerical results are of course presented
with the more realistic value
$\sin^2\theta_w=.22$.

Unless stated otherwise,
we also assume in the numerical results
a top quark mass of 175 GeV,
a collider \cm\ energy of 750 GeV,
an electron beam polarization of 90\%,
an accumulated luminosity of 20 fb$^{-1}$ and
an overall $b$- and $W$-tagging efficiency of 10\%.
The final results scale trivially like the inverse square root
of these last two collider and detector parameters,
whereas the dependence on the energy and polarization
is discussed in the text.

\section{Cross Sections and Decay}

\begin{figure}[htb]
{
\boldmath
\unitlength.5mm
\SetScale{1.418}
\begin{picture}(50,20)(-20,-110)
\ArrowLine(0,0)(15,10)
\Text(-1,0)[r]{\large$e^-$}
\ArrowLine(15,10)(0,20)
\Text(-1,20)[r]{\large$e^+$}
\Photon(15,10)(35,10){1}{4}
\Text(25,13)[cb]{\large$\gamma,Z^0$}
\ArrowLine(50,0)(35,10)
\Text(51,0)[l]{\large$\bar t$}
\ArrowLine(35,10)(50,20)
\Text(51,20)[l]{\large$t$}
\end{picture}}
\hfill
\input{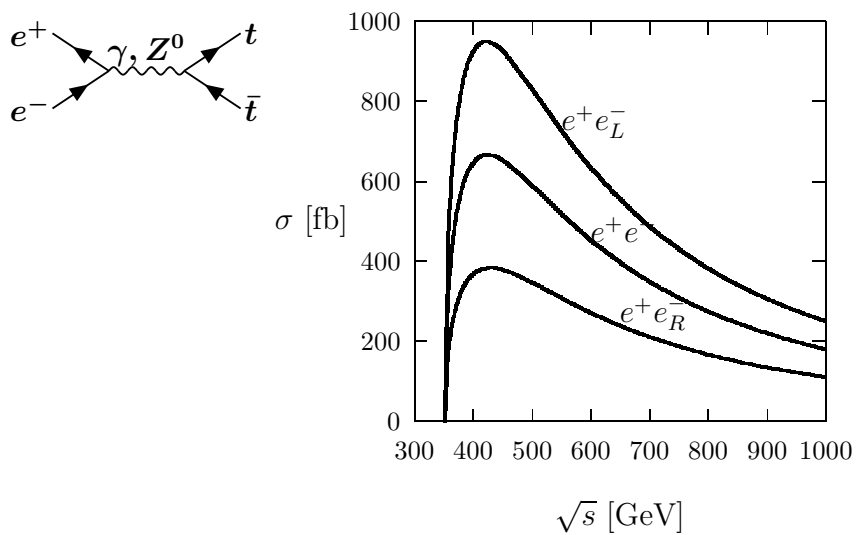}
\caption{Lowest order Feynman diagram and \xs s
for top quark pair production
in the presence and absence of polarization.
}
\label{fprod}
\end{figure}

The fermion pair-production \xs\
takes the form
\be
\sigma
\quad\simeq\quad
{A\over s^2}
\quad+\quad
{B~P\over s(s-m_Z^2)}
\quad+\quad
{C~\over (s-m_Z^2)^2}
{}~,
\label{xs}
\ee
where $A,B,C$ are constants
involving the mass and the couplings of the fermion
(here the top quark)
to the photon and the $Z^0$,
$s$ is the squared \cm\ energy of the collider
and $P$ is the polarization of the electron beam.
Clearly,
in this limit where the electron couples only
vectorially to the photon and axially to the $Z^0$,
there is no interference between the two $s$-channel
photon and $Z^0$ exchange processes
in the absence of polarization.
As can be gathered from Fig.~\ref{fprod},
for a (left) right-polarized electron beam
there are (constructive) destructive interferences.

\begin{figure}[htb]
\begin{center}
{
\boldmath
\unitlength.5mm
\SetScale{1.418}
\begin{picture}(50,20)(0,0)
\ArrowLine(0,10)(20,10)
\Text(-1,10)[r]{\large$t$}
\ArrowLine(20,10)(35,20)
\Text(36,20)[l]{\large$b$}
\Photon(20,10)(35,0){1}{4}
\Text(36,0)[l]{\large$W^+$}
\end{picture}}
\qquad\qquad\qquad
{
\boldmath
\unitlength.5mm
\SetScale{1.418}
\begin{picture}(50,40)(0,5)
\Vertex(30,15){1}
\Text(31,13)[lt]{\large$t$}
\LongArrow(30,15)(60,30)
\Text(61,30)[l]{\large$b$}
\LongArrow(30,15)(0,0)
\Text(-1,0)[r]{\large$W^+$}
\LongArrow(30,15)(30,40)
\GBox(29,15)(31,37){0}
\Text(30,41)[cb]{\large$\vec{s_t}$}
\CArc(30,15)(10,30,90)
\Text(36,26)[l]{\large$\theta$}
\end{picture}}
\end{center}
\caption{Main top decay mechanism.}
\label{fdec}
\end{figure}
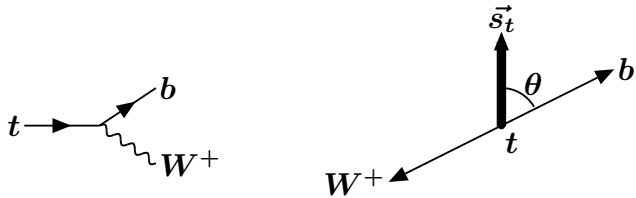

For all practical purposes
the top quark decays with a nearly 100\%\ \br\
into a bottom quark and a $W$ boson
as depicted in Fig.~\ref{fdec}.
The direction of the bottom quark
may serve as an analyzer of the top polarization.
Indeed,
in the \cm\ frame of the top,
the angle $\theta$ spanning the 3-momentum
of the bottom and the spin of the top
is distributed according to
\beq
{1\over\Gamma}{d\Gamma\over d\cos\theta}=
{1\over2} (1 \pm \beta \cos\theta)
{}~,\label{e6}
\eeq
where the mass of the bottom quark is neglected and
\beq
\beta = {m_t^2 - 2m_W^2 \over m_t^2 + 2m_W^2} \simeq {1\over3}
{}~.\label{e7}
\eeq
Measuring $CP$ violation
in the \EW\ interactions of the top quark
involves measuring the polarization of the top.
As we shall see,
the sensitivity of the measurement we propose
will thus be proportional to the polarization resolution $\beta$.

As it turns out,
if the $W$ decays leptonically,
the direction of the emerging lepton
provides an even more powerful polarization analyzer~\cite{cjk}.
In this case the resolution takes its maximum value
$\beta=1$.

\section{Top Electroweak Dipole Moments}

In the presence of \EW\ dipole moments
the interaction lagrangian of the top quark
requires the addition of the following piece

\be
{\cal L} = -{i\over2} ~{d^V} ~
\bar t \sigma^{\mu\nu}\gamma_5t~
\left(\partial_\mu V_\nu - \partial_\nu V_\mu\right)
\qquad
{V=\gamma,Z^0}
\label{dip}~,
\ee
where
$d^{\gamma,Z}$
are the electric and weak dipole moments.
The lagrangian (\ref{dip})
has three important properties:
\begin{enumerate}
\item It is not renormalizable.
  This indicates that if there is such a term,
  it cannot be an elementary interaction,
  but must originate from loop exchanges.
\item It is not $CP$ invariant.
  Therefore,
  the interactions within the aforementioned loops
  must also involve $CP$ violation.
\item It induces an helicity flip.
  The dipole moments must thus be proportional to the mass of the top quark.
\end{enumerate}

\begin{figure}[htb]
\begin{center}
{
\boldmath
\unitlength.5mm
\SetScale{1.418}
\begin{picture}(50,50)(0,0)
\Text(-6,0)[rc]{\normalsize{$t_L$}}
\ArrowLine(-5,0)(10,0)
\ArrowLine(10,0)(30,0)
\ArrowLine(30,0)(45,0)
\Text(46,0)[lc]{\normalsize{$t_R$}}
\Photon(10,0)(10,10){-1}{2}
\Text(8,5)[rc]{\normalsize{$W$}}
\Photon(30,0)(30,10){1}{2}
\Text(32,5)[lc]{\normalsize{$W$}}
\ArrowLine(30,10)(10,10)
\Text(20,8)[ct]{\normalsize{$d$}}
\ArrowLine(10,10)(20,30)
\PhotonArc(15,20)(8,65,245){1}{5.5}
\Text(7,24)[rc]{\normalsize{$W$}}
\ArrowLine(20,30)(30,10)
\Text(27,20)[lc]{\normalsize{$u$}}
\Photon(20,30)(20,40){1}{2}
\Text(20,41)[bc]{\normalsize{$V$}}
\end{picture}}
\qquad\qquad\qquad
{
\boldmath
\unitlength.5mm
\SetScale{1.418}
\begin{picture}(40,20)(0,0)
\Text(-6,0)[rc]{\normalsize{$t_L$}}
\ArrowLine(-5,0)(10,0)
\ArrowLine(10,0)(30,0)
\ArrowLine(30,0)(45,0)
\Text(46,0)[lc]{\normalsize{$t_R$}}
\DashCArc(20,0)(10,0,180){1}
\Photon(20,10)(20,20){1}{2}
\Text(20,21)[bc]{\normalsize{$V$}}
\end{picture}}
\qquad\qquad
{
\boldmath
\unitlength.5mm
\SetScale{1.418}
\begin{picture}(40,20)(0,0)
\Text(-6,0)[rc]{\normalsize{$t_L$}}
\ArrowLine(-5,0)(10,0)
\DashLine(10,0)(30,0){1}
\ArrowLine(30,0)(45,0)
\Text(46,0)[lc]{\normalsize{$t_R$}}
\ArrowArcn(20,0)(10,180,90)
\ArrowArcn(20,0)(10,90,0)
\Photon(20,10)(20,20){1}{2}
\Text(20,21)[bc]{\normalsize{$V$}}
\end{picture}}
\end{center}
\caption{Typical diagrams yielding top quark \EW\ dipole moments
in the standard model (left) and
in other theories (right).
}
\label{fdip}
\end{figure}
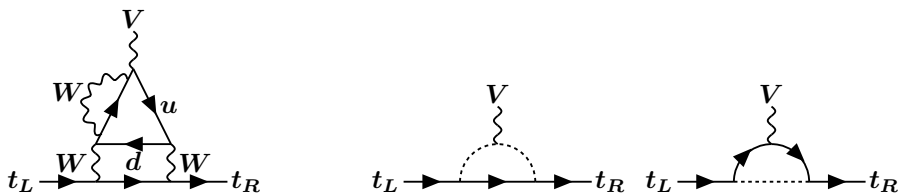

Because the dipole moments must originate from loops,
they are not constants but energy dependent form factors\cite{ff},
which develop an imaginary part beyond threshold.
At asymptotic energies
they are expected to decrease like $1/s$.

If only the complex phase of the Cabbibo-Kobayashi-Maskawa matrix is invoked
as the source of $CP$ violation,
at least three loops are necessary to generate such a term~\cite{bs},
as depicted in Fig.~\ref{fdip}.
The expected \EW\ dipole moments are therefore tiny,
around $10^{-9}$ e am
\footnote{We adopt here ``natural'' units for the dipole moments:
  the charge of the electron times atto-meters
  ($10^{-18}$ m).}.
This is far beyond the reach of any foreseeable measurement.

In theories involving non-standard sources of $CP$ violation,
however,
\EW\ dipole moments can already be generated at the one-loop level,
as depicted in Fig.~\ref{fdip}.
Su\-per\-sym\-metric, left-right symmetric and Higgs models
typically predict values
around $10^{-3}$ e am~\cite{bs}.

\section{Two Observables}

Since from the onset we expect only tiny dipole moments,
hence minute effects,
these are susceptible to be hidden by radiative corrections.
It is therefore advantageous to use $CP$-odd observables,
who to all orders have no expectation value
in the absence of \EW\ dipole moments.
We concentrate here
on the two following $CP$-odd observables~\cite{br}
\be
{\cal O}_1 = \left( \vec{p_b} \times \vec{p_{\bar b}} \right) \cdot \vec{1_z}
\qquad
\left\{\begin{array}{l}
CP \mbox{ odd} \\ CPT \mbox{ even}
\end{array}\right.
\qquad
{\Rightarrow} \mbox{ probes } {\Re e~{d}}
\ee
\be
{\cal O}_2 = \left( \vec{p_b} + \vec{p_{\bar b}} \right) \cdot \vec{1_z}
\qquad
\left\{\begin{array}{l}
CP \mbox{ odd} \\ CPT \mbox{ odd}
\end{array}\right.
\qquad
{\Rightarrow} \mbox{ probes } {\Im m~{d}}
\ee
The unit 3-vector $\vec{1_z}$
points in the same direction as the incoming positron beam.

In the presence of the $CP$-violating \EW\ dipole moments (\ref{dip}),
these observables acquire non-vanishing expectation values.
For small dipole moments
and high energies,
the latter take the approximate forms
\bea
\langle{\cal O}_1\rangle \quad \propto &
s~m_t~\beta &
\left[
  A~\Re e {d^\gamma} ~P
  +
  B~\Re e {d^Z}
\right]
\label{ev1}
\\
\langle{\cal O}_2\rangle \quad \propto &
\sqrt{s}~m_t~\beta~ &
\left[
  C~\Im m {d^\gamma}
  +
  D~\Im m {d^Z} ~P
\right]
\label{ev2}~,
\eea
where $A \dots D$ are complicated constants.
As anticipated,
these expectation values are proportional
to the polarization resolution $\beta\simeq.33$.
As we alluded to earlier,
observables involving the momentum of the decay leptons of the $W$'s
instead of the bottom quark
would have an improved resolution $\beta=1$.
The obvious trade-off is the inevitable loss in statistics~\cite{pr}.

Although Eqs~(\ref{ev1},\ref{ev2}) are only approximations,
they show that in the absence of a polarized electron beam
little information can be gained about
the real part of the electric dipole moment and
the imaginary part of the weak dipole moment.

At this stage it is worth mentioning
that since the positron beam cannot be polarized,
the initial state is not a $CP$ eigenstate.
Therefore $CP$-odd correlations are not
necessarily a measure of the $CP$ violation of the interaction.
If we neglect the electron mass and radiative corrections,
though,
only the left-right and right-left combinations of electron and positron
helicities couple to the photon and $Z^0$.
This makes the effectively contributing intial state
indeed a $CP$ eigenstate.

Nevertheless,
even in the limit of vanishing electron mass
hard collinear photons can flip the helicities
of the initial electrons or positrons~\cite{fs}.
This can lead to non-zero $CP$-odd
correlations even in the absence of $CP$-violating interactions.
However,
since ${\cal O}_1$ is $T$-odd,
this $CP$-conserving hard photon emission mechanism
can only contribute to $\langle{\cal O}_1\rangle$
if its amplitude has an absorptive part,
{\em i.e.},
at higher order.
This argument does not hold for ${\cal O}_2$
which is $T$-even.
The \sm\ contribution to $\langle O_2\rangle$,
though,
has been shown to be negligible~\cite{ar}
compared to the variance
$\langle{\cal O}_2^2\rangle$
({\em cf.} next section).
If one insists,
this background correlation can of course also be subtracted
or removed by a simple cut on the total energy of the event.

\section{Discovery Limits}

\begin{figure}[htb]
\centerline{\input{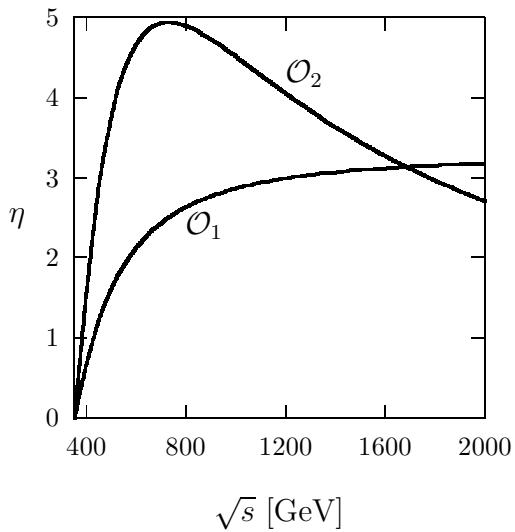}}
\caption{Energy dependence of the number of standard deviations
in Eq.~(\protect\ref{ss}),
with 20 fb$^{-1}$ of data
with 90\%\ right polarized beams,
if $d^Z=0$ and $d^\gamma=0.1+i0.5$ e am.
}
\label{feny}
\end{figure}

To be statistically significant,
the expectation values (\ref{ev1},\ref{ev2})
must be larger than the expected natural variances
of the observables $\langle{\cal O}^2\rangle$.
A signal of $\eta$ standard deviations
is obtained for a sample of $N$ events
if
\be
\langle{\cal O}\rangle \geq \eta
{}~\sqrt{\langle{\cal O}^2\rangle \over N}
\label{ss}~.
\ee
The analytical expressions for these variances~\cite{cr}
are long and not particularly enlightning.
We therefore do not present them here.
Let us only comment
that asymptotically they are proportional to
the squared \cm\ energy $s$.
Therefore
the resolving power of the observable ${\cal O}_1$
saturates,
whereas the resolving power of ${\cal O}_2$
eventually decreases at high energies.
As can be gathered from Fig.~\ref{feny},
the best accuracy is obtained around 750 GeV.

\vskip-5mm

This having been said,
one must bear in mind
that the dipole moments are merely form factors.
Since the value of their real part decreases with the collider energy,
there should also be an optimum energy
for the observable ${\cal O}_1$.
For the sake of concreteness
we perform the rest of this analysis
with a \cm\ energy of 750 GeV.

\begin{figure}[htb]
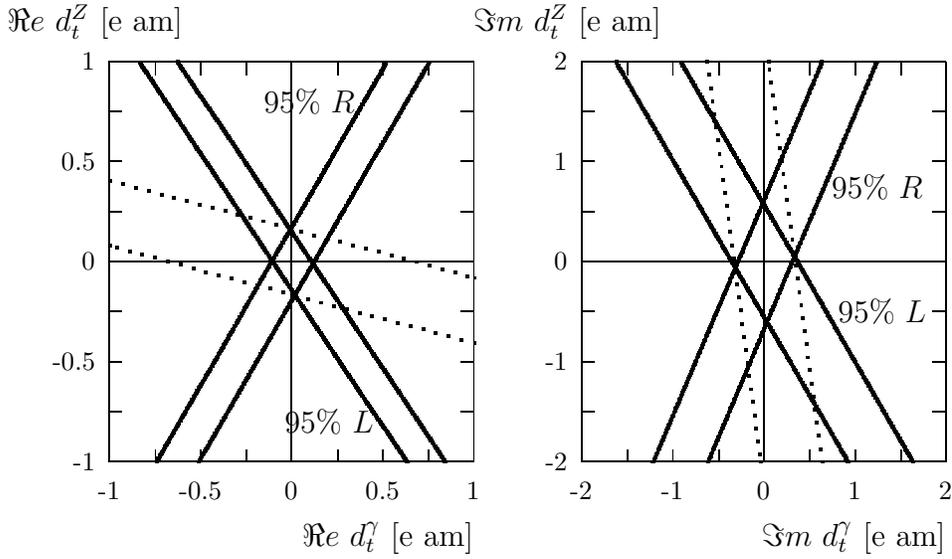

\input{re.tex}
\input{im.tex}
\caption{Bounds on the real and imaginary parts
of the top electric and weak dipole moments,
to be obtained at 750 GeV with 20 fb$^{-1}$ of data.
The dotted lines are for an unpolarized electron beam.
}
\label{freim}
\end{figure}

In Figs~\ref{freim}
we display the areas in the $(d_t^\gamma,d_t^Z)$ plane
which cannot be explored to better than 3 standard deviations
in Eq.~(\ref{ss}).
Because of the linear dependence
of the expectation values (\ref{ev1},\ref{ev2}) on the dipole moments,
these areas are straight bands
centered around the \sm\ expectation
$d_t^\gamma = d_t^Z = 0$.
The slopes of these bands
vary with the polarization of the initial electron beam.
For fully polarized beams they are narrowest.
As the degree of polarization is decreased,
they rotate around fixed points and become wider.

As anticipated from the approximate Eqs~(\ref{ev1},\ref{ev2}),
Figs~\ref{freim} show that in the absence of polarization
the observables ${\cal O}_1$ and ${\cal O}_2$
are rather insensitive to ${\cal R}e~d^{\gamma}_t$ and ${\cal I}m~d^{Z^0}_t$
respectively.
Moreover,
a single measurement
with or without polarization
cannot exclude large dipole moments:
in some unfortunate situations,
the electric and weak dipoles
can assume large values,
while their effects cancel out
so that no $CP$ violation is apparent.
However,
if the information from two measurements
with opposite electron polarization is combined,
both the electric and weak dipoles
can be constrained simultaneously
down to values around $10^{-1}$ e am.

\section{Conclusions}

Because of its large mass,
the top quark is a privileged candidate
for carrying electric and weak dipole moments.
Still,
the \sm\ predicts such tiny values for these dipole moments,
that any observation thereof
would be a ``gold-plated'' indication of new physics.

We have analyzed two $CP$-odd observables
in top pair-production and decay
and have come to the conclusion
that their resolving power
is substantially enhanced
in the presence of polarized electron beams.

We find that \EW\ dipole moments
down to $10^{-1}$ e am
can be probed.
This is still about two orders magnitude larger
than what is usually expected
from most theories extending beyond the \sm.
However,
we are confident
that a lot more can be gained
by combining the information also gathered
from other (as or more efficient~\cite{as})
observables
in polarized $e^+e^-$ and $\gamma\gamma$ scattering.

\section*{Acknowledgments}
It is a pleasure to thank Saurabh Rindani
for his collaboration in this project.

\end{document}